# Superconductivity in Ir-doped $LaFe_{1-x}Ir_xAsO$

Yanpeng Qi, Lei Wang, Zhaoshun Gao, Dongliang Wang, Xianping Zhang, Zhiyu Zhang, Yanwei Ma*

Key Laboratory of Applied Superconductivity, Institute of Electrical Engineering, Chinese Academy of Sciences, P. O. Box 2703, Beijing 100190, China

**Abstract:**

We report the realization of superconductivity by 5d element Ir doping in LaFeAsO, a prototype parent compound of high-temperature iron based superconductors. X-ray diffraction patterns indicate that the material has formed the ZrCuSiAs–type structure with a space group P4/nmm. The systematic evolution of the lattice constants demonstrated that the Fe ions were successfully replaced by the Ir. Both electrical resistance and magnetization measurements show superconductivity up to 11.8 K in $LaFe_{1-x}Ir_xAsO$. The superconducting transitions at different magnetic fields were also measured yielding a slope of $-dH_{c2} / dT = 6.7$ T / K near Tc. Using the Werthamer-Helfand-Hohenberg (WHH) formula $H_{c2} = -0.69 (dH_{c2} / dT)|_{Tc} Tc$, the upper critical field at zero K is found to be about 54 T. Hall effect measurements indicate that the conduction in this material is dominated by electron-like charge carriers, the charge carrier density determined at 100 K is about $3.71 \times 10^{20} / cm^3$, which is close to the $LaFeAsO_{1-x}F_x$ system. This is the first example of bulk superconductivity induced by replacing the Fe sites with higher d-orbital electrons in FeAs-1111 family, which should add more ingredients to the underlying physics of the iron-based superconductors.

* Author to whom correspondence should be addressed; E-mail: ywma@mail.iee.ac.cn

## 1. Introduction

Until recently the chemical realm of high-Tc superconductivity had been limited mainly to copper oxide-based layered perovskites, and now much attention has been focused on novel Fe-based superconductors since the recent discovery of superconductivity at the high temperature of 26 K in $LaFeAsO_{1-x}F_x$ [1]. The parent compounds LaFeAsO (denoted as FeAs-1111) have a quasi two-dimensional tetragonal structure, consisting of insulating La-O layers and conducting FeAs layers. Similar to the cuprates, the Fe-As layer is thought to be responsible for superconductivity and La-O layer is carrier reservoir layer to provide electron carrier. By replacing La with other rare earth elements, $T_c$ can be raised to above 50 K [2-7]. Thus, a new class of materials with a promising potential for high Tc that may rival the well-known cuprate high-temperature superconductors was born. Besides, the oxygen-free iron arsenide compounds $AFe_2As_2$ (denoted as FeAs-122, where $A$ = Ba, Sr, Ca, Eu) were discovered and superconductivity was found by appropriate substitution or under pressure [8-13]. Also other compounds like LiFeAs [14] and FeSe [15] were found to exhibit superconductivity, even at lower temperatures, thus opening the way to a new "iron-age".

The undoped compound LaFeAsO itself is not superconducting but shows an anomaly at around 150 K, electron doping by F suppresses the anomaly and recovers the superconductivity [1]. It is widely accepted that chemical doping has become a very important strategy to induce superconductivity in the iron oxyarsenides compounds. In contrast to high-Tc cuprates, superconductivity can also be induced by partial substitution of Fe by other 3d transition mental elements like Co and Ni in the iron based compounds [16-18]. We consider that the detailed investigation of doping in the FeAs planes would be helpful to shed light on the mechanisms of these new superconductors, so it is worthwhile to research the substitution at Fe site with different elements, however, the early experiments were focused on the replacement of the Fe site with ones nearby the iron with 3d orbital electrons. Very recently, 4d or 5d transition metals such Ru and Ir substitution at Fe sites in the FeAs-122 phase have also been exhibited superconductivity [19-21]. Therefore, it is intriguing to know whether it is possible to induce superconductivity by substituting Fe ions with other transition metals in the FeAs-1111 compounds. In this work, we report the observation of bulk superconductivity in $LaFe_{1-x}Ir_xAsO$. X-ray diffraction (XRD), SEM-EDX, resistivity, DC magnetic susceptibility as well as Hall effect measurements have been performed for the system of $LaFe_{1-x}Ir_xAsO$. Our results demonstrate that superconductivity can be realized by replacing the Fe sites with higher d-orbital electrons not only in the FeAs-122 family, but also in the FeAs-1111 phase.

## 2. Experimental

The polycrystalline $LaFe_{1-x}Ir_xAsO$ samples were synthesized by using a one-step solid state reaction method developed by our group. The details of fabrication process are described elsewhere [6, 18]. Stoichiometric amounts of the starting elements La, Fe, $Fe_2O_3$, Ir and As were thoroughly grounded and encased into pure Nb tubes. After packing, this tube was subsequently rotary swaged and sealed in a Nb tube. The sealed samples were heated to 1180 $^oC$ and kept at this temperature for 45 hours. It is noted that the sample preparation process except for annealing was performed in glove box in which high pure argon atmosphere is filled. Then it was cooled down slowly to room temperature. The high purity argon gas was allowed to flow into the furnace during the heat-treatment process.

The x-ray diffraction measurement was performed at room temperature using an MXP18A-HF-type diffractometer with Cu- $K_\alpha$ radiation from 20° to 80° with a step of 0.01°. The analysis of x-ray powder diffraction data was done and the lattice constants were derived. Microstructure observations were performed using scanning electron microscope (SEM/EDX). The DC magnetization measurements were done with a superconducting quantum interference device (SQUID). The zero- field- cooled magnetization was measured by cooling the sample at zero field to 2 K, then magnetic field was applied and the data were collected during the warming up process. The field-cooled magnetization data was collected in the warming up process after the sample was cooled down to 2 K at a finite magnetic field. The resistance and Hall effect measurements were done by using a physical property measurement system (Quantum Design, PPMS) with magnetic fields up to 9 T .

**3. Results and discussion**

Figure 1 shows the x-ray diffraction patterns of $LaFe_{1-x}Ir_xAsO$. It is seen that all main peaks can be well indexed based on the ZrCuSiAs tetragonal structure with the space group P4/nmm, indicting that the samples are nearly single phase. There are still some small peaks coming from the second phase, as marked by the asterisks. Further analysis indicates that this tiny amount of impurity is most probably the LaAs. By fitting the data to the structure calculated with the software X'Pert Plus, we got the lattice constants. In figure 2, we show *a*-axis and *c*-axis lattice parameters for the $LaFe_{1-x}Ir_xAsO$ samples. The lattice parameters of undoped sample are $a$ = 4.0328 Å, $c$ = 8.7372 Å, however, Ir-doping leads to an apparent decrease in *c*-axis lattice while the *a*-axis increases a bit. Similar behavior is observed in the $SrFe_{2-x}Ir_xAs_2$ compounds [21]. Compared to the parent compound LaFeAsO, the apparent variation of the lattice parameters upon Ir-doping indicates a successful chemical substitution in $LaFe_{1-x}Ir_xAsO$ compounds.

Figure 3 shows the scanning electron microscope images of samples with the nominal formula $LaFe_{0.925}Ir_{0.075}AsO$. As one can see, the samples seem denser though there are several voids observed. In addition, multiple layers forming a large grain of the superconducting phase can be easily detected in the samples, as shown in Fig. 3 (b), indicating a layer growth mechanism. Typical EDX spectrum of $LaFe_{0.925}Ir_{0.075}AsO$ samples is presented in Fig. 4. The analyzed results are also shown in Table I. Clearly, EDX analysis of the grains revealed the presence of uniformly distributed La, As, Fe, Ir, and O, suggesting that superconducting grains are compositionally homogeneous, at least within the limits of SEM-EDX analysis. In particular, our EDX data make evident that the Ir is really entered into the lattice, and the true doping levels are close to the nominal ones in the most cases. It should be noted that the same was found for other nominal compositions of samples.

Figure 5 shows the temperature dependence of the electrical resistivity for $LaFe_{1-x}Ir_xAsO$ samples in the temperature range from 2 to 300 K. The inset shows an enlarged plot of $\rho$ versus T at the low temperatures. It is clear that the undoped

LaFeAsO sample exhibits a clear anomaly near 150 K [1], which is ascribed to the spin-density-wave (SDW) instability and structural phase transitions from tetragonal to orthorhombic symmetry. By doping more Ir, the resistivity drop was converted to an uprising. As seen from the Fig. 5, the electrical resistivity of the x = 0.05 increases obviously below 80 K, which is similar to undoped sample, and finally we can observe a rapid transition to zero resistivity. This is similar to the case of Co doping in the iron oxyarsenides compounds [22, 23]. At higher Ir-doping, the uprising in the lower temperature is not obvious, and the highest transition temperature 11.8 K is observed at x = 0.075. The transition width is ~ 2 K, indicating a good quality of our samples. With further doping the transition temperature declines slightly. The superconductivity disappeared again when the doping content x is 0.20. Our results suggest that superconductivity can be easily induced in the FeAs-1111 family by replacing the Fe sites, regardless of the transition metals of 3d or higher d-orbital elements. It should be noted that the absolute value of resistivity derived from our polycrystalline samples here may be larger than that of single crystals due to the grain boundary scattering and the porosity (see Fig.3).

To further confirm the superconductivity of $LaFe_{1-x}Ir_xAsO$, DC magnetic susceptibility measurement was also performed. Figure 6 shows the temperature dependence of DC magnetization for the $LaFe_{1-x}Ir_xAsO$ (x = 0.05, 0.075, 0.10) samples. The measurements were carried out under a magnetic field of 20 Oe in zero-field-cooled and field-cooled processes. Strong diamagnetic signals can be seen around 11 K for x = 0.075, which is corresponding well to the middle transition point of resistance; a superconducting volume fraction is large enough to constitute bulk superconductivity. For the samples of x = 0.05 and 0.1, the diamagnetic signals are not so strong. The inset of Fig. 6 shows the rough superconducting phase diagram in $LaFe_{1-x}Ir_xAsO$. A dome-like Tc (x) curve can be seen, which is similar to other iron based superconductors [1-5].

Figure 7 shows the temperature dependence of resistivity for $LaFe_{0.925}Ir_{0.075}AsO$ under different magnetic fields. Similar to other iron based superconductors, applied magnetic field is observed to suppress the transition. It is noted that the onset transition temperature is not sensitive to magnetic field, but the zero resistance point shifts more quickly to lower temperatures due to the weak links or flux flow, which is similar to that of $AFe_2As_2$ single crystal. We tried to estimate the upper critical field ($H_{c2}$) and irreversibility field ($H_{irr}$), using the 90% and 10% points on the resistive transition curves. The change of transition temperature ($T_c$) with critical field (H) is shown in the inset of Fig. 7. It is clear that the curve of $H_{c2}$ (T) is very steep with a slope of $- dH_{c2}/dT|_{Tc} = 6.7$ T / K. This value is obviously larger than that obtained in $SrFe_{2-x}Ir_xAs_2$ compounds [21]. From this figure, using the Werthamer-Helfand-Hohenberg formula [24], $H_{c2}(0) = 0.693 \times (dH_{c2} / dT) \times T_c$, we can get $H_{c2}(0) \approx 54.3$ T. If adopting a criterion of 99 $\%\rho_n(T)$ instead of $90\%\rho_n(T)$, the $H_{c2}(0)$ value of this sample obtained by this equation is even higher.

In order to get more information about the conducting carriers in the samples, we also carried out the Hall effect measurements. Figure 8 shows the Hall resistivity $\rho_{xy}$ for the sample x = 0.075. It is clear that all the curves have good linearity versus the magnetic field and $\rho_{xy}$ is negative at all temperatures above the critical temperature, indicating that the normal state conduction of $LaFe_{0.925}Ir_{0.075}AsO$ is dominated by the electron-like charge carriers. From this set of data, the Hall coefficient $R_H = \rho_{xy} /H$ is determined and shown in Fig.9. One can see that the Hall coefficient changes slightly at high temperatures but drops below 100 K. It is known that the Hall coefficient is a constant versus temperature for a normal metal with the Fermi liquid feature; however, this situation is changed for a multiband material or a sample with non-Fermi liquid behavior, such as the curate superconductors. The temperature dependence of Hall coefficient indicates that either the mulitband effect or some unusual scattering process may be involved in our samples. If using the single band equation $n = 1 /R_H e$ to evaluate the charge carrier density at 100 K, we could obtain $n = 3.71 \times 10^{20}$ /cm$^3$, which is similar to the value of $LaFeAsO_{1-x}F_x$ [25]. It should be noted that both system, $LaFe_{1-x}Ir_xAsO$ and $LaFeAsO_{1-x}F_x$, have a low charge carrier density. This would give support to a theoretical proposal that the iron-based superconductors have very low superfluid density [26].

The layered transition metal oxides attract a great deal of attention in superconductivity community since the discovery of high-temperature superconductivity in cuprates. The continual search for new superconductors has led to the discovery of superconductivity in 4d-transition metal ruthenate $Sr_2RuO_4$ (Tc ≈ 1.4 K) [27], and 3d-transition metal cobaltate $Na_xCoO_2 \cdot H_2O$ (x < 0.35, y < 1.3) (Tc ≈ 4 K) [28], however, their transition temperatures are much lower than the copper oxides. The new iron based superconductors REFeAsO is another layered transition metal oxides, which is the first non-cuprate high-Tc superconductors with the transition temperature higher than 40 K. In addition, it is still unclear why the superconducting transition temperature varies in those layered transition metal oxides. It is widely believed that the high Tc values of the copper oxides are related to the strong electron correlation associated with the transition metal ions, and recently theoretical calculations demonstrated that 3d orbitals of Fe atoms contribute to the multiple Fermi surfaces, hence superconductivity in iron based compounds [29]. The substitution of Fe site with other d-band element not only opens up new possibilities for exploring novel superconducting compounds, but also offers opportunity to study the origin of superconductivity from transition metal d-band electron. Our data indicate that superconductivity could be induced in the FeAs-1111 family by replacing the Fe sites, regardless of the transition metals of 3d or higher d-orbital electrons, which will add extra ingredients in understanding the underlying physics in the iron based superconductors.

**4. Conclusions**

In summary, we have fabricated the new superconductor $LaFe_{1-x}Ir_xAsO$ with a maximum Tc about 11.8 K by replacing the Fe with the 5d-transition metal Ir. The presence of zero resistance and diamagnetism in the measurement proves that Ir substitution in the LaFeAsO compounds leads to superconductivity. The superconductivity is rather robust against the magnetic field with a slope of $-dH_{c2} / dT = 6.7$ T / K near Tc. Using the Werthamer-Helfand-Hohenberg (WHH) formula $H_{c2} = -0.69 (dH_{c2} / dT)|_{Tc} Tc$, the upper critical field at zero K is found to be about 54 T. The Hall coefficient is negative indicating the electrical transport behavior. The charge carrier density at 100 K is about $3.71 \times 10^{20} / cm^3$, which is close to the $LaFeAsO_{1-x}F_x$ system. Our results suggest that superconductivity can be realized in FeAs-1111 compounds by replacing the Fe sites with different transition metal elements which are not restricted to ones nearby the iron with 3d orbital electrons.

**Acknowledgments**

The authors thank Profs. Haihu Wen, Liye Xiao and Liangzhen Lin for their help and useful discussions. This work was partially supported by the Beijing Municipal Science and Technology Commission under Grant No. Z09010300820907, National Science Foundation of China (grant no. 50802093) and the National ‘973’ Program (grant no. 2006CB601004).

## References

[1] Y. Kamihara, T. Watanabe, M. Hirano and H. Hosono, J. Am. Chem. Soc. **130**, 3296 (2008).

[2] X. H. Chen, T. Wu, G. Wu, R. H. Liu, H. Chen and D. F. Fang, Nature **453**, 376 (2008).

[3] H. H. Wen, G. Mu, L. Fang, H. Yang and X. Zhu, Europhys. Lett. **82**, 17009 (2008).

[4] G. F. Chen, Z. Li, D. Wu, G. Li, W. Z. Hu, J. Dong, P. Zheng, J. L. Luo and N. L. Wang, Phys. Rev. Lett. **100**, 247002 (2008).

[5] Z. A. Ren, J. Yang, W. Lu, W. Yi, X. L. Shen, Z. C. Li, G. C. Che, X. L. Dong, L. L. Sun, F. Zhou and Z. X. Zhao, Chin. Phys. Lett. **25**, 2215 (2008).

[6] Y. W. Ma, Z. S. Gao, L. Wang, Y. P. Qi, D. L. Wang, X. P. Zhang, Chin. Phys. Lett. 26, 037401 (2009).

[7] L. Wang, Z. S. Gao, Y. P. Qi, D. L. Wang, X. P. Zhang and Y. W. Ma, Supercond. Sci. Technol. **22**, 015019 (2009).

[8] M. Rotter, M. Tegel and D. Johrendt, Phys. Rev. Lett. **101**, 107006 (2008).

[9] K. Sasmal, B. Lv, B. Lorenz, A. Guloy, F. Chen, Y. Xue and C. W. Chu, Phys. Rev. Lett. **101**, 107007 (2008).

[10] Z. S. Wang, H. Q. Luo, C. Ren and H. H. W, Phys. Rev. B **78**, 140501(R) (2008)

[11] H. S. Jeevan, Z. Hossain, D. Kasinathan, H. Rosner, C. Geibel and P. Gegenwart, Phys. Rev. B. **78**, 092406 (2008).

[12] Y. P. Qi, Z. S. Gao, L. Wang, D. L. Wang, X. P. Zhang and Y. W. Ma, New J. Phys. **10,** 123003 (2008).

[13] M. S. Torikachvili, S. L. Bud'ko, N. Ni and P. C. Canfield, Phys. Rev. Lett., **101**, 057006 (2008).

[14] X.C.Wang, Q.Q. Liu, Y.X. Lv, W.B. Gao, L.X.Yang, R.C.Yu, F.Y.Li, C.Q. Jin, arXiv: 0806.4688

[15] Fong-Chi Hsu, Jiu-Yong Luo, Kuo-Wei Yeh, Ta-Kun Chen, Tzu-Wen Huang, Phillip M. Wu, Yong-Chi Lee, Yi-Lin Huang, Yan-Yi Chu, Der-Chung Yan,

Maw-Kuen Wu, PNAS **105**, 14262 (2008).

[16] A. S. Sefat, R. Jin, M. A. McGuire, B. C. Sales, D. J. Singh, and D. Mandrus, Phys. Rev. Lett. 101, 117004 (2008).

[17] L. J. Li, Q. B. Wang, Y. K. Luo, H. Chen, Q. Tao, Y. K. Li, X. Lin, M. He, Z. W. Zhu, G. H. Cao and Z. A. Xu, New J. Phys. **11**, 025008 (2009).

[18] Y. P. Qi, Z. S. Gao, L. Wang, D. L. Wang, X. P. Zhang and Y. W. Ma, Supercond. Sci. Technol. **21**, 115016 (2008).

[19] S. Paulraj, S. Sharma, A. Bharathi, A. T. Satya, S. Chandra, Y. Hariharan and C. S. Sundar, Cond-mat: arXiv, 0902.2728 (2009).

[20] Y. P. Qi, Z. S. Gao, L. Wang, D. L. Wang, X. P. Zhang and Y. W. Ma, Cond-mat: arXiv, 0903.4967 (2009).

[21] F. Han, X. Zhu, Y. Jia, L. Fang, P. Cheng, H. Luo, B. Shen and H. H. Wen, Cond-mat: arXiv, 0902.3957 (2009).

[22] A. S. Sefat, A. Huq, M. A. McGuire, R. Y. Jin, B. C. Sales, D. Mandrus, L. M. D. Cranswick, P. W. Stephens and K. H. Stone, Phys. Rev. B **78**, 104505 (2008).

[23] G. H. Cao, C. Wang, Z. W. Zhu, S. Jiang, Y. K. Luo, S. Chi, Z. Ren, Q. Tao, Y. T. Wang and Z. A. Xu, Phys. Rev. B **79**, 054521 (2009).

[24] N. R. Werthamer, E. Helfand and P. C. Hohenberg, Phys. Rev. **147**, 295-302 (1966)

[25] X. Y. Zhu, H. Yang, L. Fang, G. Mu and H. H. Wen, Supercond. Sci. Technol. **21** 105001 (2008).

[26] D. J. Singh and M. H. Du, Phys. Rev. Lett. **100**, 237003 (2008)

[27] Y. Maeno, H. Hashimoto, K. Yoshida, S. Nishizaki, T. Fujita, J. G.Bednorz, F. Lichtenberg, Nature **372**, 532 (1994).

[28] K. Takada, H. Sakurai, E. Takayama-Muromachi, F. Izumi, R. A. Dilanian and T. Sasaki, Nature **422**, 53 (2003).

[29] H. Ding, P. Richard, K. Nakayama, K. Sugawara, T. Arakane, Y. Sekiba, A. Takayama, S. Souma, T. Sato, T. Takahashi, Z. Wang, X. Dai, Z. Fang, G. F. Chen, J. L. Luo and N. L. Wang, Eur. Phys. Lett. **83,** 47001 (2008).

## Captions

Figure 1 XRD patterns of the $LaFe_{1-x}Ir_xAsO$ samples. The impurity phases are marked by *.

Figure 2 Doping dependence of *a*- and *c*-axis lattice constant. It is clear that the *a*-axis lattice expands, while *c*-axis lattice shrinks with the Ir substitution.

Figure 3 (a) Low magnification and (b) high magnification SEM micrographs for the $LaFe_{0.925}Ir_{0.075}AsO$ samples.

Figure 4 The Energy dispersive X-ray microanalysis (EDX) spectrums of the $LaFe_{0.925}Ir_{0.075}AsO$ samples. The little rectangles (S1-3) represent the positions where we performed the EDX analysis.

Figure 5 Temperature dependence of resistivity for the $LaFe_{1-x}Ir_xAsO$. Inset: Enlarged view of low temperature, showing superconducting transition.

Figure 6 Temperature dependence of DC magnetization for the $LaFe_{1-x}Ir_xAsO$ (x = 0.05, 0.075, 0.10) samples. The measurement was done under a magnetic field of 20 Oe in the zero-field-cooled and field-cooled modes. Inset: the superconducting phase diagram in $LaFe_{1-x}Ir_xAsO$.

Figure 7 Temperature dependence of resistivity for $LaFe_{0.925}Ir_{0.075}AsO$ samples at different magnetic fields. Inset: The upper critical field $H_{c2}$ and $H_{irr}$ as a function of temperature for $LaFe_{0.925}Ir_{0.075}AsO$ samples.

Figure 8 Hall resistivity with relation of magnetic field at different temperatures for $LaFe_{0.925}Ir_{0.075}AsO$ samples.

Figure 9 Temperature dependence of Hall coefficient for $LaFe_{0.925}Ir_{0.075}AsO$ samples, the negative value indicates that the charge carrier is electron type.

Table I Atomic ratio of the elements for the $LaFe_{0.925}Ir_{0.075}AsO$ samples.

| **Element** | **O** | **Fe** | **As** | **La** | **Ir** |
|---|---|---|---|---|---|
| **S1** | 26.63 | 25.07 | 24.32 | 22.59 | 1.39 |
| **S2** | 24.76 | 25.74 | 24.60 | 23.32 | 1.59 |
| **S3** | 25.77 | 25.69 | 24.02 | 23.13 | 1.40 |

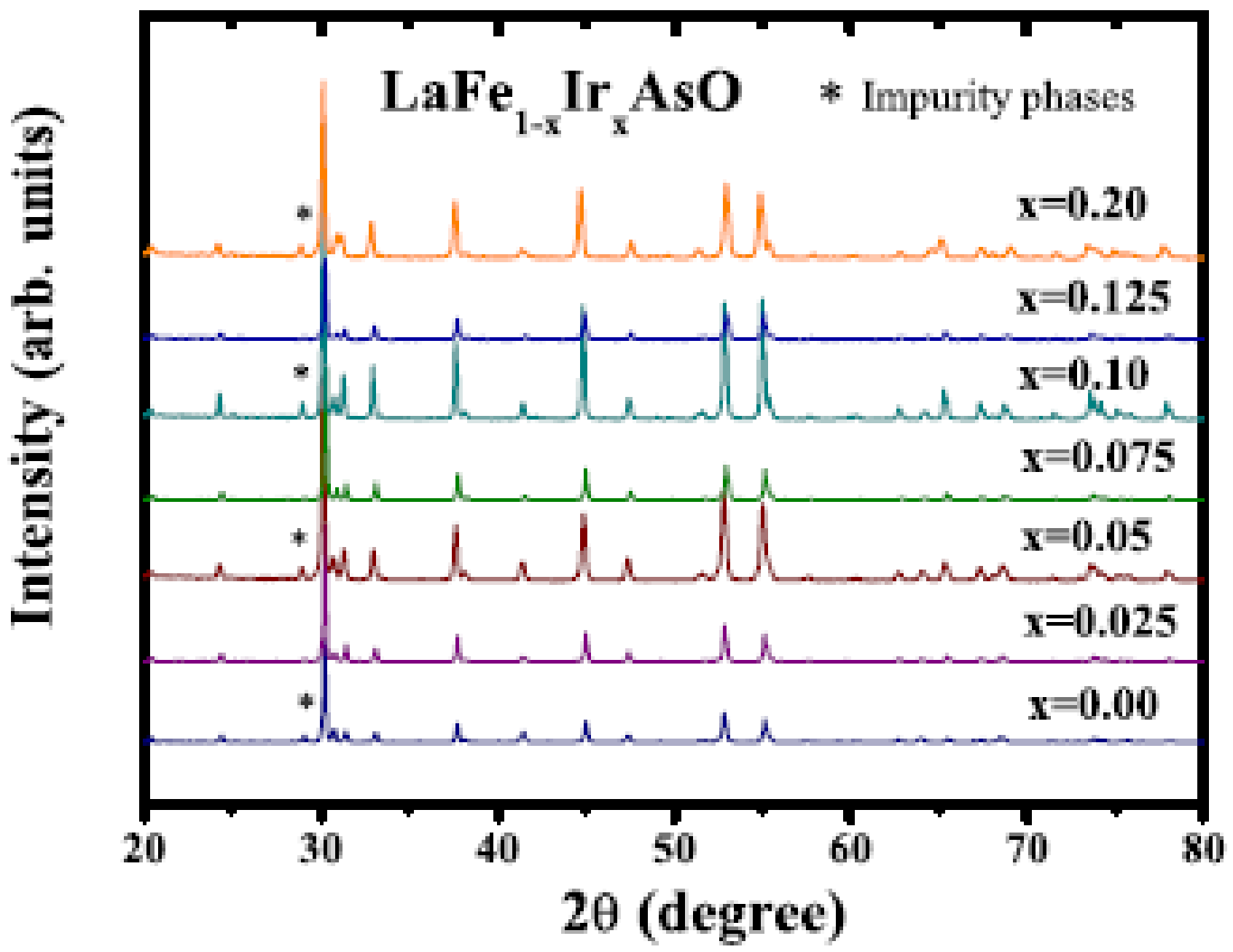

Fig.1 Qi et al.

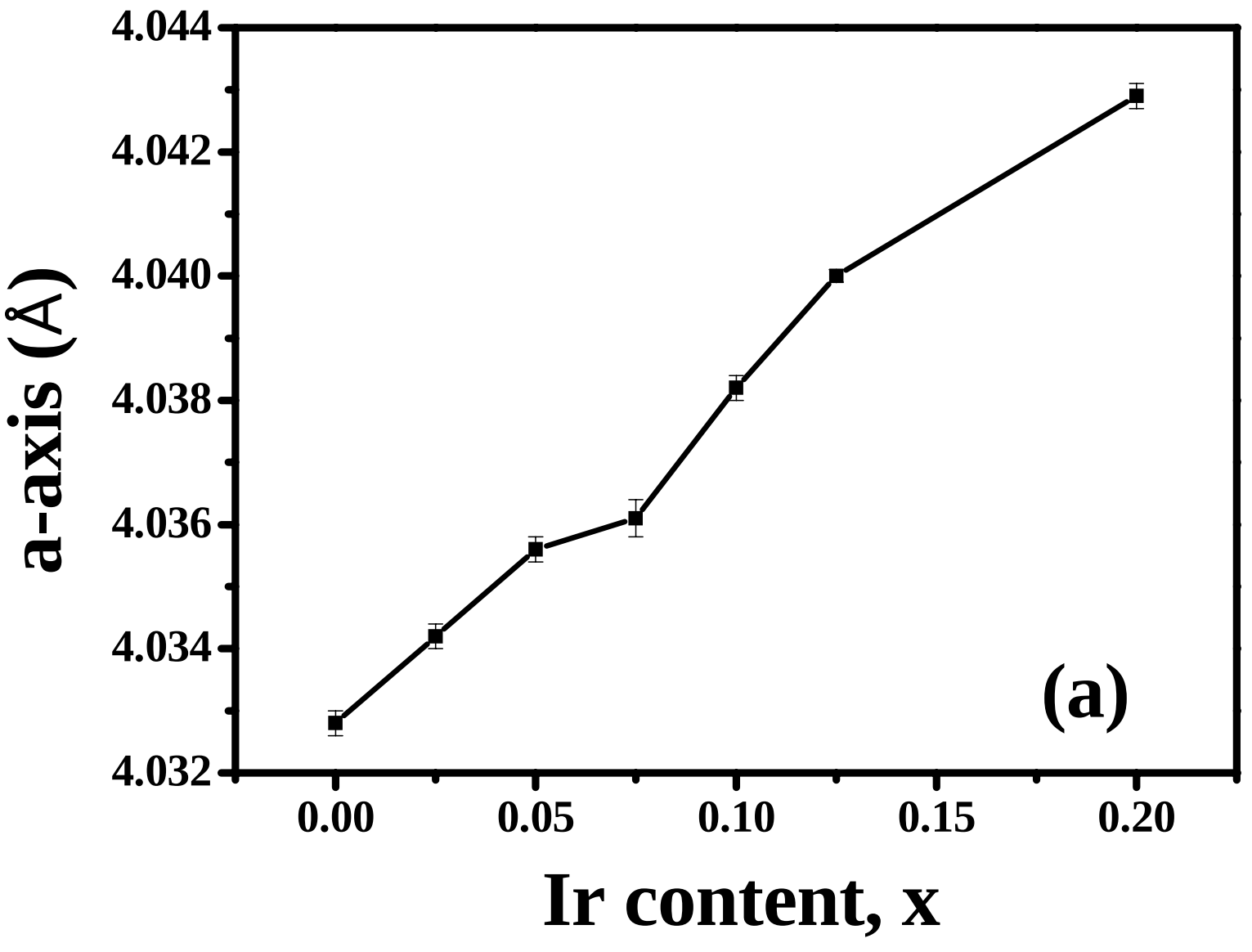

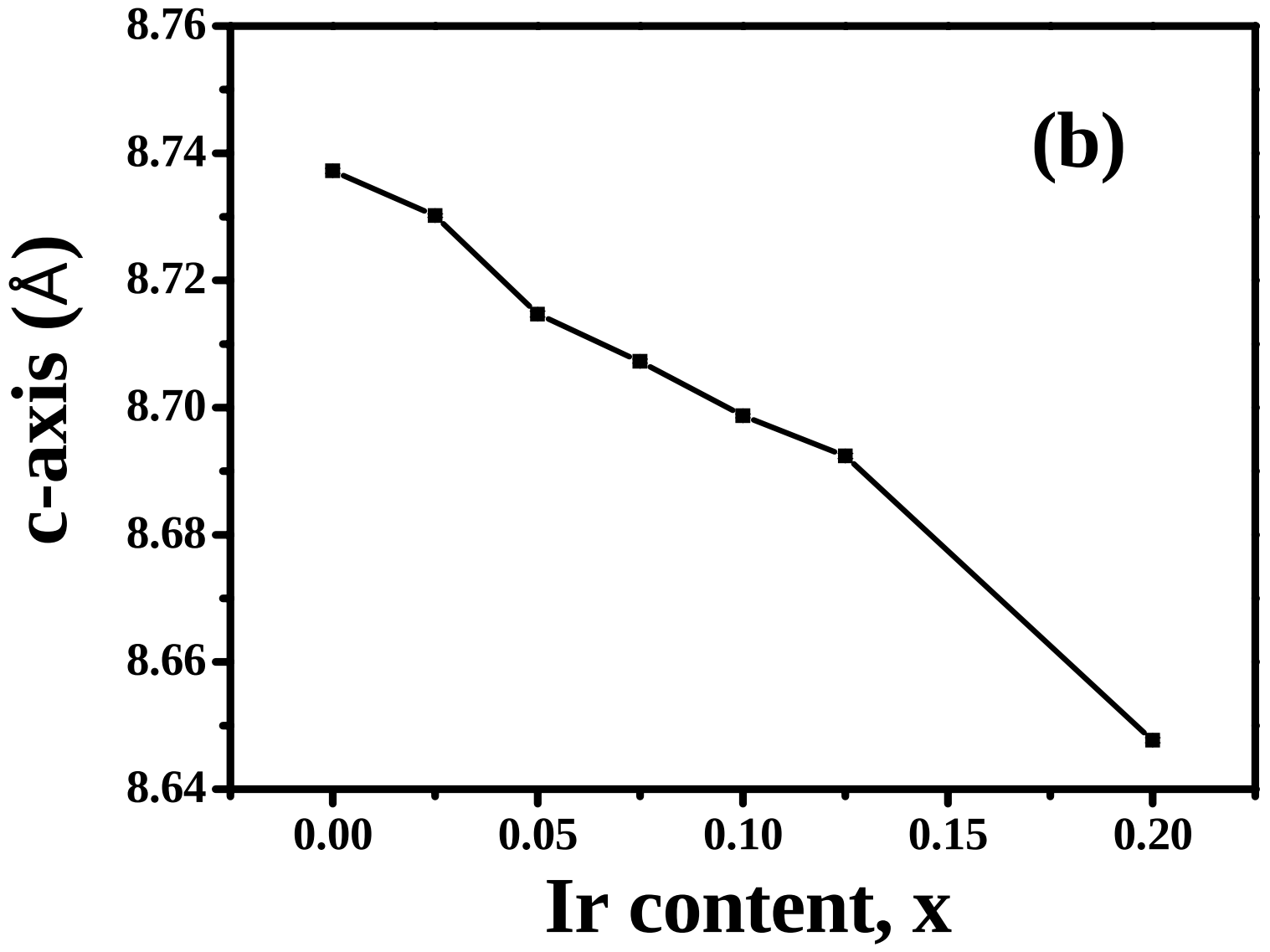

Fig.2 Qi et al.

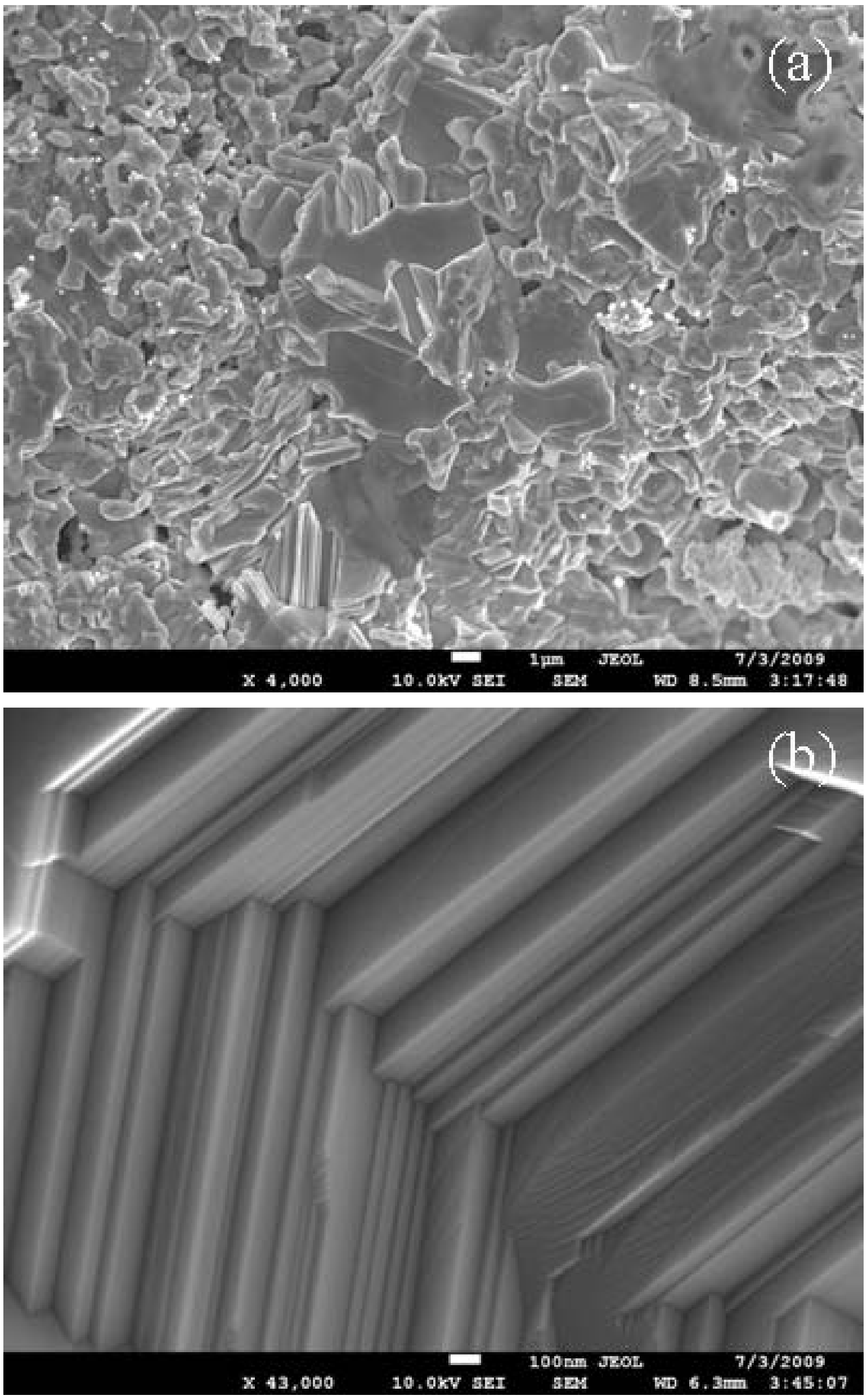
(a)
1μm JEOL 7/3/2009
X 4,000 10.0kV SEI SEM WD 8.5mm 3:17:48
(b)
100nm JEOL 7/3/2009
X 43,000 10.0kV SEI SEM WD 6.3mm 3:45:07

Fig.3 Qi et al.

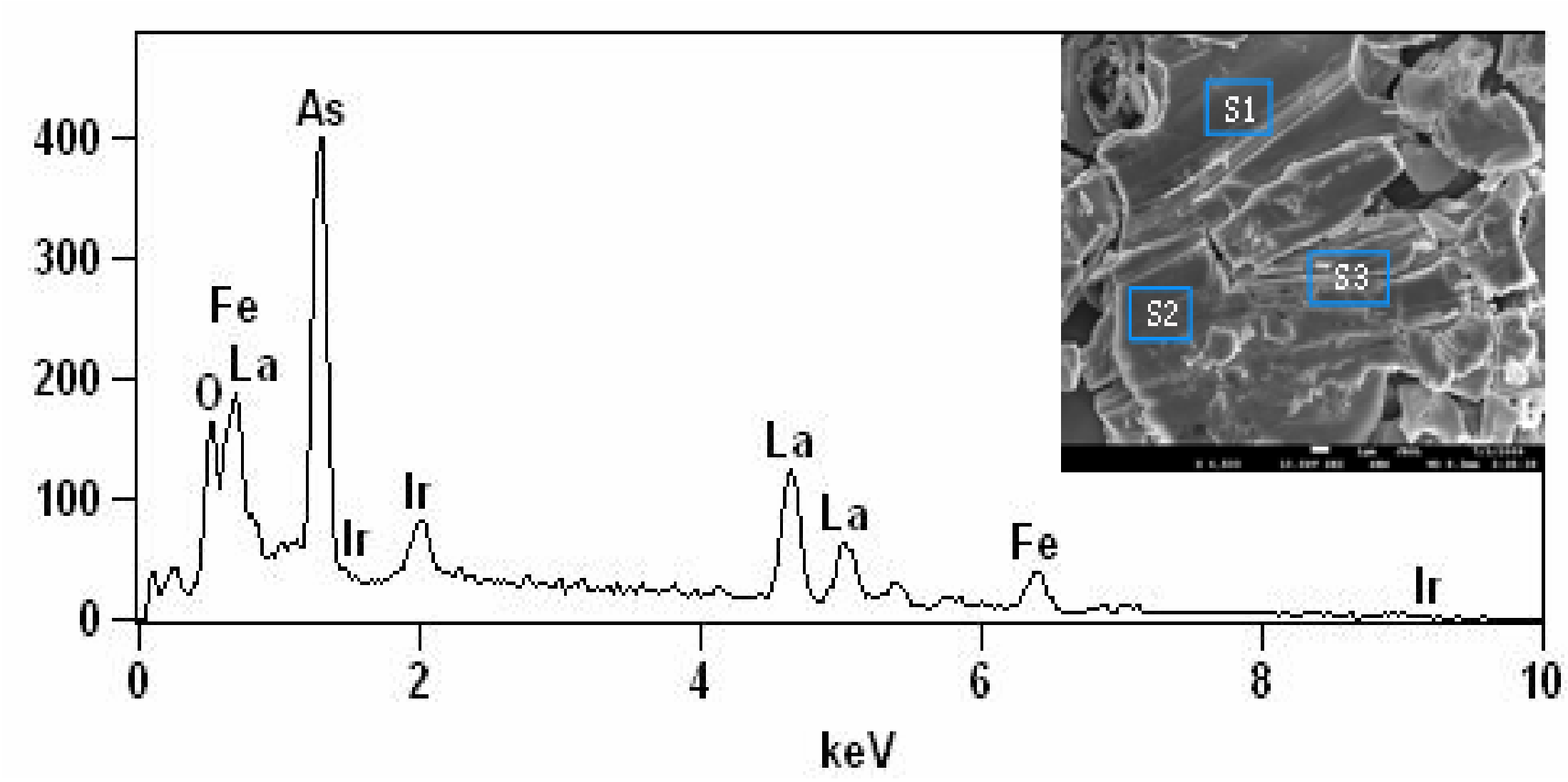

Fig.4 Qi et al.

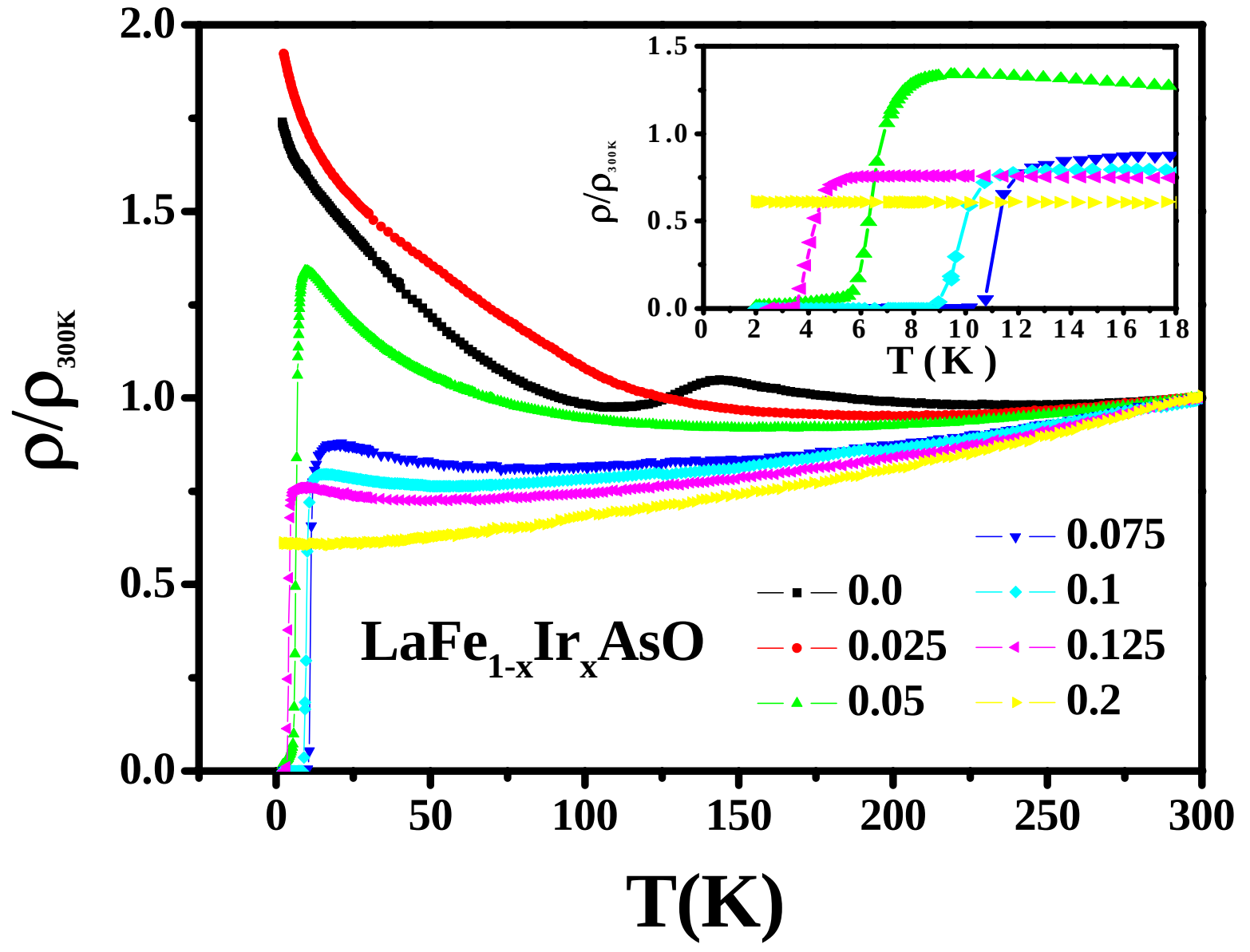

Fig.5 Qi et al.

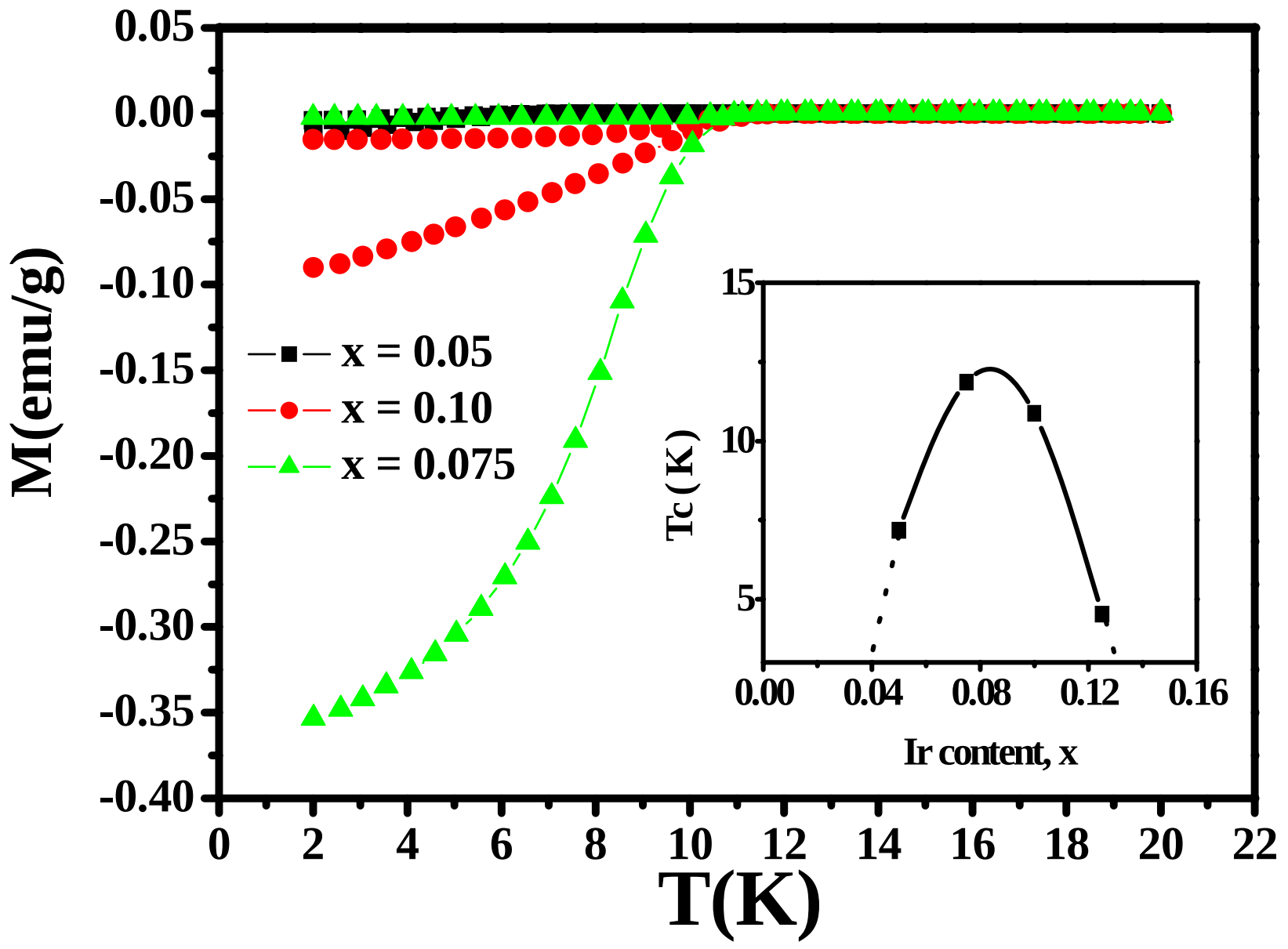

Fig.6 Qi et al.

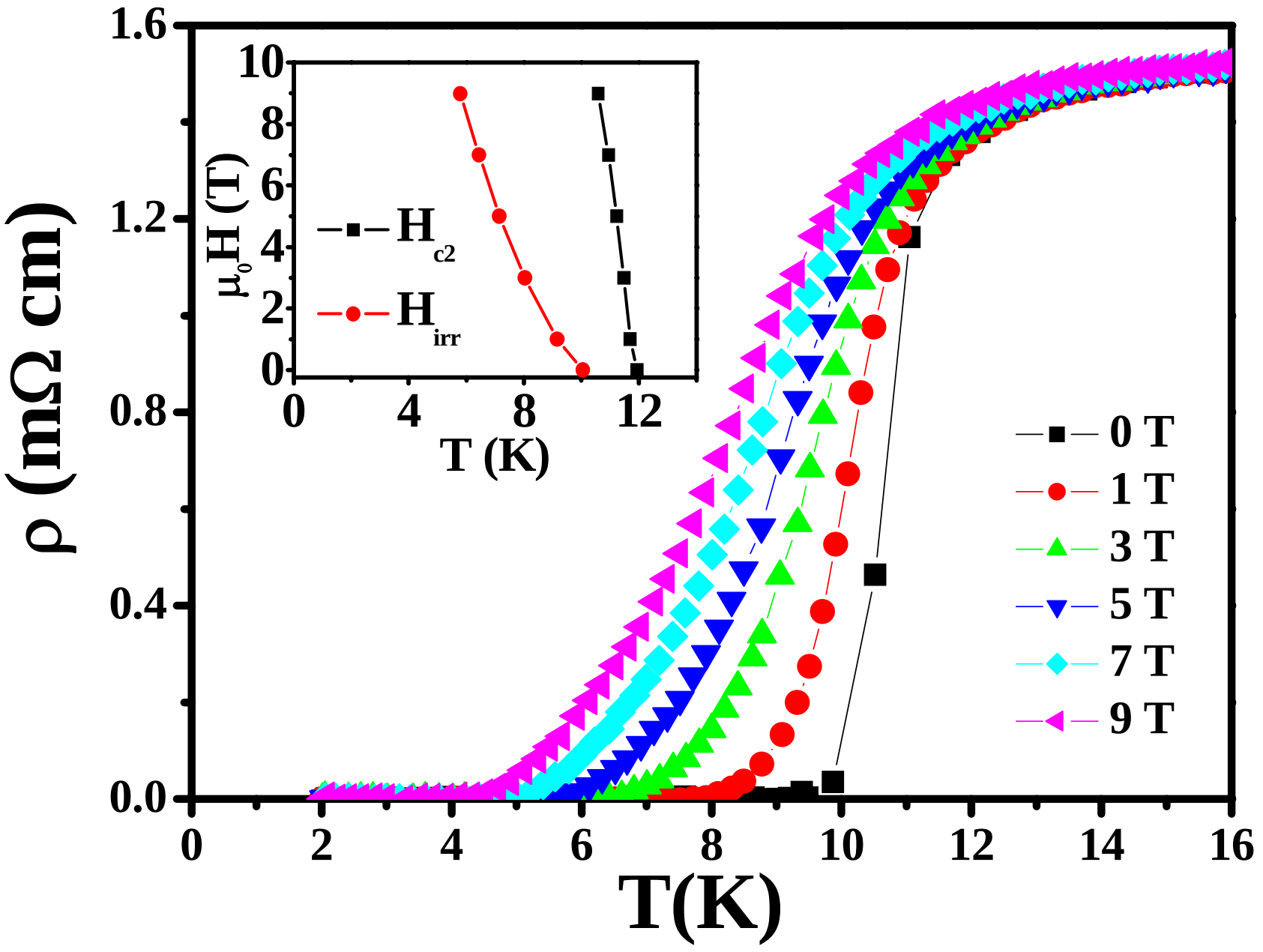

Fig.7 Qi et al.

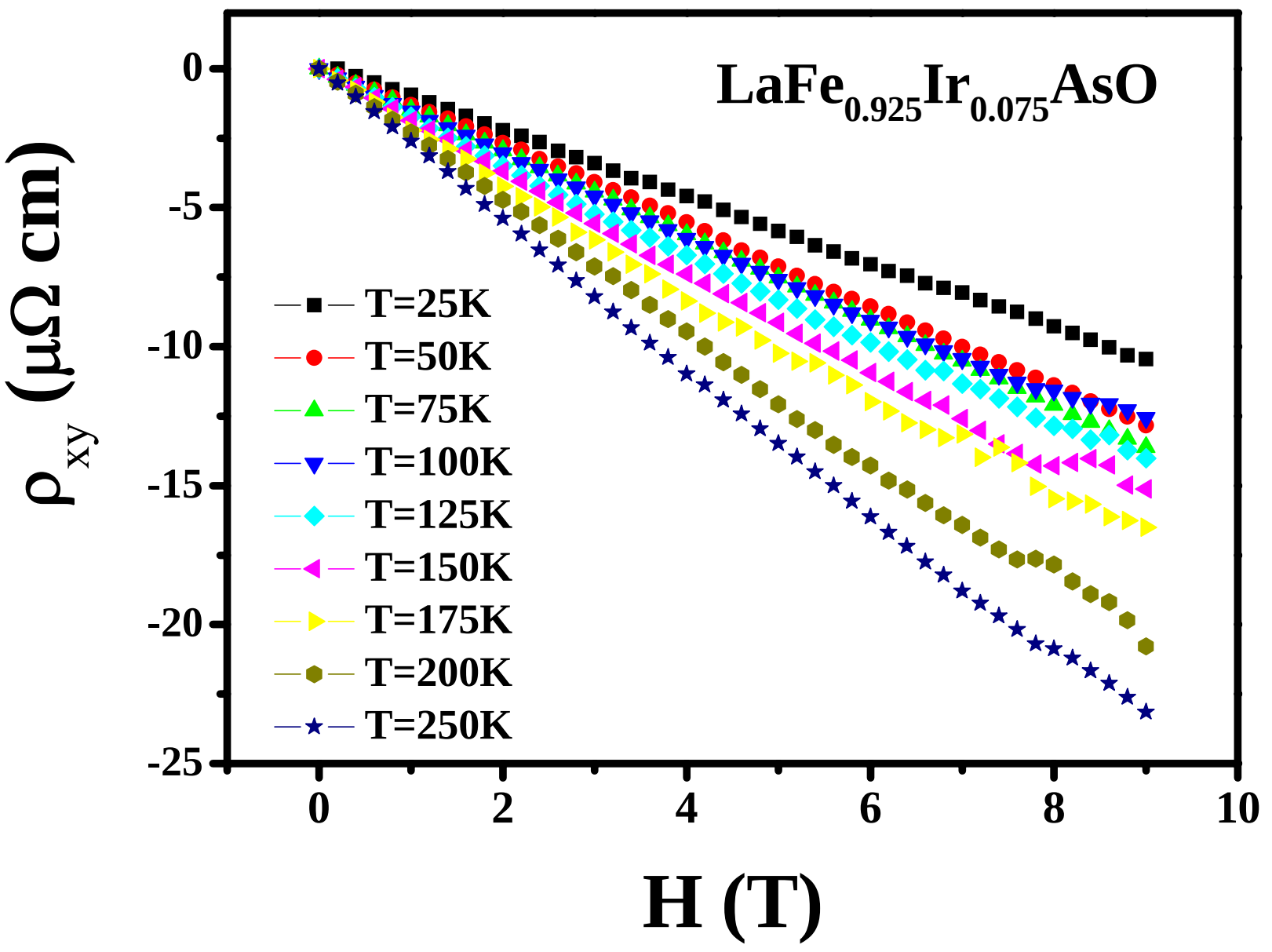

Fig.8 Qi et al.

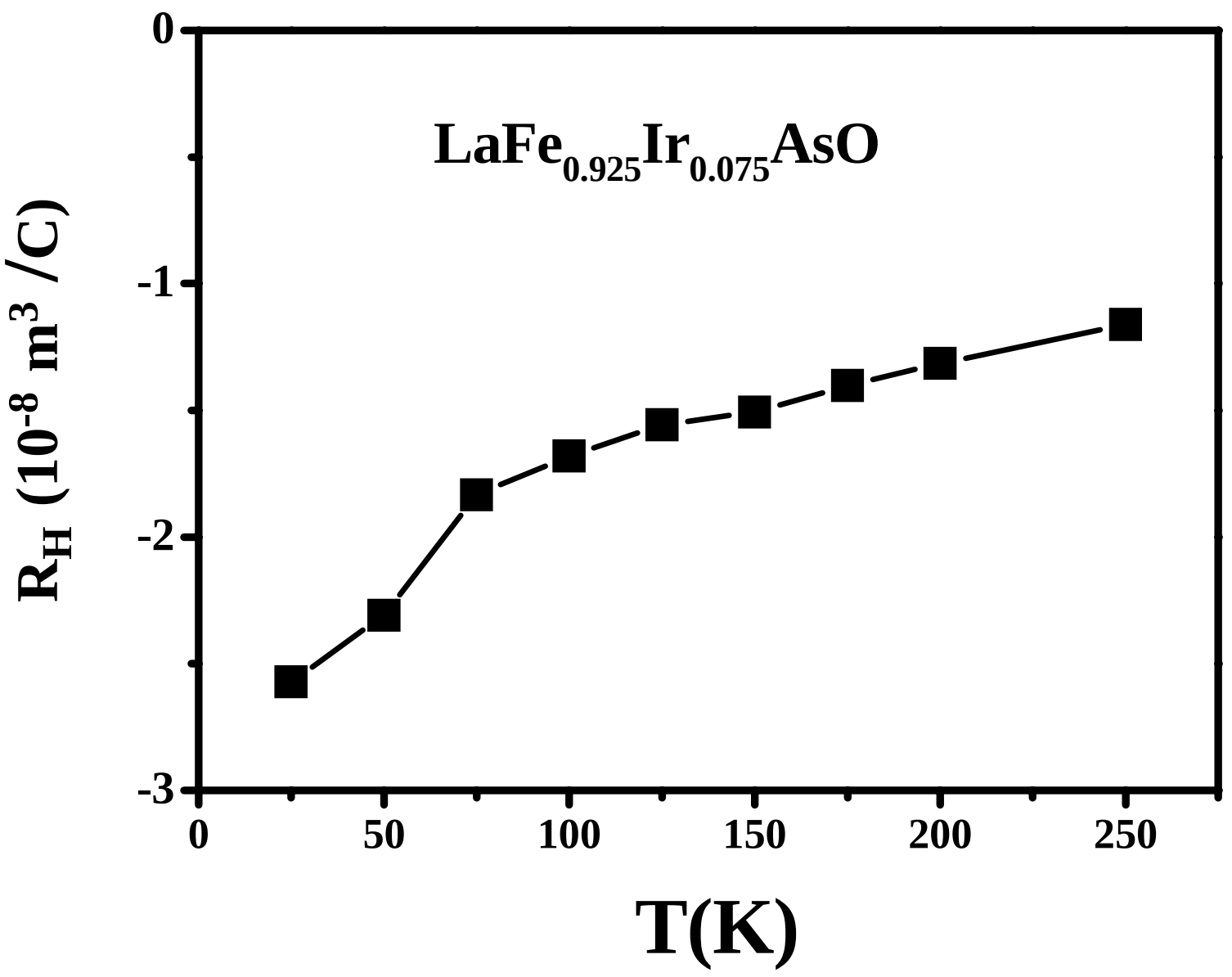

Fig.9 Qi et al.